\documentclass[aps,prl,reprint]{revtex4-1}

\usepackage[utf8]{inputenc}

\usepackage{hyperref}

\usepackage{xcolor}
\usepackage{makecell}
\usepackage{graphicx}

\usepackage{amsmath}

\newcommand{\delG}{\delta_G}
\newcommand{\fG}{f_G}
\newcommand{\aG}{a_G}
\newcommand{\bG}{b_G}
\newcommand{\cG}{c_G}
\newcommand{\dG}{d_G}
\newcommand{\delD}{\delta_D}
\newcommand{\fD}{f_D}
\newcommand{\aD}{a_D}
\newcommand{\bD}{b_D}
\newcommand{\cD}{c_D}
\newcommand{\dD}{d_D}

\newcommand{\Jdc}{J_\text{dc}}
\newcommand{\X}{\mathbf{X}}
\newcommand{\Xdot}{\dot{\mathbf{X}}}
\newcommand{\FST}{\mathbf{F}^\text{ST}}
\newcommand{\ez}{\mathbf{e}_z}
\newcommand{\OM}{\Bar{\Bar{\Omega}}}
\newcommand{\Ms}{M_\text{s}}
\newcommand{\lex}{l_\text{ex}}
\newcommand{\HOe}{\mathbf{H}_\text{Oe}}

\newcommand{\kST}{\kappa^\text{ST}_\perp}
\newcommand{\kex}{k^\text{ex}}
\newcommand{\kOe}{\kappa^\text{Oe}}
\newcommand{\kms}{k^\text{ms}}

\newcommand{\Cp}{C^+}
\newcommand{\noOe}{noOF}
\newcommand{\Cm}{C^-}

\newcommand{\titre}{Data-driven Thiele equation approach for solving the full nonlinear spin-torque vortex oscillator dynamics}

\begin{document}

\title{\titre}
\author{Flavio ABREU ARAUJO}
\email{flavio.abreuaraujo@uclouvain.be}
\affiliation{Institute of Condensed Matter and Nanosciences, Universit\'{e} catholique de Louvain, Place Croix du Sud 1, 1348 Louvain-la-Neuve, Belgium}
\author{Chloé CHOPIN}
\affiliation{Institute of Condensed Matter and Nanosciences, Universit\'{e} catholique de Louvain, Place Croix du Sud 1, 1348 Louvain-la-Neuve, Belgium}
\author{Simon DE WERGIFOSSE}
\affiliation{Institute of Condensed Matter and Nanosciences, Universit\'{e} catholique de Louvain, Place Croix du Sud 1, 1348 Louvain-la-Neuve, Belgium}

\begin{abstract}
The dynamics of vortex based spin-torque nano-oscillators is investigated theoretically. Starting from a fully analytical model based on the Thiele equation approach, fine-tuned data-driven corrections are carried out to the gyrotropic and damping terms. These adjustments, based on micromagnetic simulation results, allow to quantitatively model the response of such oscillators to any dc current within the range of the vortex stability. Both, the transient and the steady-state regimes are accurately predicted under the proposed data-driven Thiele equation approach. Furthermore, the computation time required to solve the dynamics of such system is reduced by about six orders of magnitude compared to the most powerful micromagnetic simulations. This major breakthrough opens the path for unprecedented high-throughput simulations of spin-torque vortex oscillators submitted to long-duration input signals, for example in neuromorphic computing applications.   
\end{abstract}

\maketitle

\section{Introduction}

Artificial neural network (ANN) algorithms have emerged as one of the most successful machine learning paradigms in recent years \cite{hopfield1988artificial,thrun1994learning,jain1996artificial,silver2017mastering,jumper2021highly}. However typical software implementation of ANNs are generally energy-consuming, driving the seek for low-power solutions. In this framework, neuromorphic computing aims to draw inspiration from the human brain architecture to propose energy efficient hardware implementations of ANNs capable of solving complex cognitive tasks. Different physical devices mimicking neuronal properties have been considered up to now, such as memristors \cite{milano2021materia} or transistors \cite{roy2019towards}. All are characterized by a nonlinear response to a given stimulus, analogously to their biological equivalent. Among other implementations explored, spin-torque vortex oscillators (STVOs) present many advantages, such as nanoscale, low-noise and great tunability \cite{pribiag2007magnetic}. These devices, based on magnetic tunnel junctions (MTJs) in which the free ferromagnetic layer hosts a non-uniform magnetic vortex state, have already shown convincing results in various speech recognition applications \cite{torrejon2017neuromorphic,romera2018vowel}.\\
Different approaches exist for modeling the dynamics of magnetic vortices. Among these, the micromagnetics formalism is well-recognized to give quantitative results comparable to experimental measurements \cite{leliaert2019tomorrow}. Based on a continuum approximation, it consists in solving time-dependent equations of magnetization dynamics. Despite the accuracy of their results, micromagnetic simulations (MMS) present the main disadvantage of requiring important computational power as well as being time-consuming. This makes them ineffective for handling any signal encountered in practical applications. An alternative is to use analytical equations, much faster to solve, describing the vortex motion. For example, very simple models (i.e., exponential decay) have already been used to approximate the vortex core displacement in transient regime of oscillation \cite{araujo2020role}. More commonly, the Thiele equation approach \cite{thiele1973steady} (TEA) is examined. Albeit based on physical considerations, it has though given poor results compared to simulations up until now because of cumbersome mathematical derivations involved for a vortex topology.

In this work, we propose a new hybrid method (i.e., semi-analytical) which is both fast and accurate to describe the vortex core precessions. Based on a small set of simulations which have reached steady-state oscillating regime, we can adapt the gyrotropic and damping terms appearing in the Thiele equation. This allows to capture the STVO response to any current of arbitrary form in transient and steady-state oscillating regime. We show that this technique can accelerate the computation time by several orders of magnitude compared to micromagnetism, while remaining as precise.

\section{Methods}
The dynamics of vortices confined inside ferromagnetic nanopillars can be described using the Thiele equation approach \cite{thiele1973steady}. The case of a free layer of radius $R$ and thickness $h$ will be considered hereafter. Within this framework, the vortex core is seen as a quasi-particle and identified by its in-plane position $\mathbf{X}=(X,Y)$ inside the cylindrical dot. Following the TEA, the evolution of the vortex core position is given as 
\begin{equation}
    G(\ez \times \Xdot) + D \Xdot = \frac{\partial W}{\partial \X} + \FST,
    \label{eq:thiele}
\end{equation}
where $G$ and $D$ are the gyrotropic and damping constants, respectively, $W$ is the potential magnetic energy associated to a displacement of the vortex core and $\FST$ are the forces related to the spin-transfer torque. In a previous work \cite{araujo2022ampere} we showed that, considering a perpendicular polarizer, Eq.~(\ref{eq:thiele}) could be reduced to the following system of linear first-order differential equations
\begin{equation}
    \left[ {\begin{array}{c}
    \Dot{X} \\
    \Dot{Y} \\
    \end{array} } \right] =
  \underbrace{\left[ {\begin{array}{cc}
    \Gamma & -\omega \\
    \omega &  \Gamma \\
  \end{array} } \right]}_{\OM}
  \left[ {\begin{array}{c}
   X \\
   Y \\
  \end{array} } \right],
  \label{eq:matrix}
\end{equation}
where the parameters $\Gamma$ and $\omega$ are given as
\begin{align}
    \Gamma &= \frac{D(\kex + \kms + C\Jdc\kOe) + G\Jdc\kST}{D^2 + G^2},\\
    \omega &= \frac{D\Jdc\kST - G(\kex + \kms + C\Jdc\kOe)}{D^2 + G^2},
\end{align}
where $C=\pm 1$ is the vortex chirality and $\Jdc$ the direct current density. The other terms are explicited hereafter.

In this simple harmonic oscillator model, we considered three contributions to $W$, namely the exchange, magneto-static and Ampère-Oersted field energies. Those are expressed using their spring-like restoring force constants $k$ as
\begin{equation}
     \frac{\partial W}{\partial \X} = (\kex + \kms + \underbrace{C \Jdc \kOe}_{k^\text{Oe}})\X.
\end{equation}
Here and below, the notation $s=\sqrt{X^2 + Y^2}/R$ will be used to refer to the vortex core reduced position. The exchange energy contribution writes as \cite{guslienko2001field,gaididei2010magnetic}
\begin{equation}
    \kex = (2\pi)^2 h \Ms^2 (\lex/R)^2 / (1-s^2),
    \label{eq:kex}
\end{equation}
where $\Ms$ is the saturation magnetization and $\lex=\sqrt{A/(2\pi\Ms^2)}$ is the exchange length of the material, with $A$ the exchange stiffness coefficient. The magneto-static component is given as \cite{gaididei2010magnetic,araujo2022ampere}
\begin{equation}
    \kms_\xi = \frac{8\Ms^2 h^2}{R}\textstyle \Lambda_{0,\xi}\left(1+a_\xi s^2+b_\xi s^4 + c_\xi s^6\right),
\end{equation}
where $\xi=h/(2R)$ corresponds to the aspect ratio of the nanodot. The parameters $\Lambda_{0,\xi}$, $a_\xi$, $b_\xi$ and $c_\xi$ can be calculated thanks to numerical methods we described in \cite{araujo2022ampere}. Finally, the contribution related to the Ampère-Oersted field has been derived as follows \cite{araujo2022ampere}
\begin{equation}
    \kOe = \frac{8\pi^2}{75}\Ms R h \left(1-\frac{4}{7}s^2 - \frac{1}{7} s^4 - \frac{16}{231} s^6 - \frac{125}{3003}s^8 \right).
\end{equation}
In the absence of any external magnetic field, the latter accounts for the entire Zeeman energy.

As stated previously, we fixed the polarization direction to be out-of-plane. This means that only the perpendicular component of the Slonczewski spin-torque \cite{slonczewski1996current} contributes to the STVO dynamics. We have thus $\FST=\kST \Jdc (\mathbf{e}_z \times \X)$ with \cite{khvalkovskiy2009vortex,dussaux2012field}
\begin{equation}
    \kST = \pi a_J \Ms h p_z,
\end{equation}
where $\mathbf{p}=p_z\mathbf{e}_z=(0,0,1)$ is the unit vector giving the polarization direction of the fixed layer, $a_J=p_J\hbar/(2|e|\Ms^\text{ref}h)$ is the spin-transfer efficiency with $p_J$ the spin-current polarization, $e$ is the electron charge and $\Ms^\text{ref}$ the polarizer saturation magnetization.

In our previous model, we supposed that the gyro- and damping constants were independent of the vortex core position. We had thus $G=G_0=-2\pi P\Ms h/\gamma_\text{G}$ \cite{guslienko2002eigenfrequencies,guslienko2006low}, where $P$ is the vortex polarity and $\gamma_\text{G}=g|e|/(2m_e)$ is the gyromagnetic ratio with $g$ the electron spin $g$-factor and $m_e$ the electron mass, and $D=D_0=-\alpha_\text{G} \eta |G|$ with $\alpha_\text{G}$ the Gilbert damping constant and $\eta =\ln(R/(2\lex))/2+3/8$ \cite{khvalkovskiy2009vortex}. This assumption led to quantitative predictions in the resonant regime (i.e., $s=0$) but lack of precision for nonlinear auto-oscillations. In the present paper we propose to take the $s$-dependence of $G(s)$ and $D(s)$ into account.

As the calculation of the latter is a very difficult task, no convincing expression exists in the literature, to the best of our knowledge, to describe their evolution with respect to the core position. Thence, we chose a more convenient approach. The gyro- and damping terms are modeled as $G(s) = G_0\fG(s)$ and $D(s) = D_0 \fD(s)$, respectively. The functions $\fG(s)$ and $\fD(s)$ are supposed to be even power expansions of $s$, so that
\begin{align}
    \fG(s) &= 1 + \aG s^2 + \bG s^4 + \cG s^6 + \dG s^8 \label{eq:fG},\\
    \fD(s) &= 1 + \aD s^2 + \bD s^4 + \cD s^6 + \dD s^8 \label{eq:fD}.
\end{align}
This definition allows to retrieve $G(0)=G_0$ and $D(0)=D_0$, as $\fG(0)=1$ and $\fD(0)=1$. In addition, the coefficients appearing in Eqs~(\ref{eq:fG})~\&~(\ref{eq:fD}) can be deduced from a limited set of simulations. By fitting micromagnetic results to our analytical model, it is thus possible to considerably enhance its predictions for $s\neq 0$. This step is performed by following the procedure described hereafter.

Using the matrix $\OM$ from Eq.~(\ref{eq:matrix}), the following system is derived
\begin{equation}
\begin{cases}
    D(s)\Gamma - G(s) \omega = \kex(s) + \kms(s) + C \Jdc \kOe(s)\\
    D(s)\omega + G(s) \Gamma = \Jdc \kST
\end{cases}
\end{equation}
As $\Gamma=0$ in steady-state oscillating regime \cite{araujo2022ampere} and using the expressions of $G(s)$ and $D(s)$, one finds

\begin{align}
    \delG\fG(s) &= -\frac{ \kex(s) + \kms(s) + C \Jdc \kOe(s)}{G_0 \omega(s)} \label{eq:dfG},\\
    \delD\fD(s) &= \frac{\Jdc \kST}{D_0 \omega(s)} \label{eq:dfD},
\end{align}
where we introduced two global correction factors $\delG$ and $\delD$. Those aim to absorb any imprecision of the model resulting from the multiple assumptions made, even at $s=0$. Those discrepancies could originate from the analytical expression of $G_0$ and $D_0$ as well as from the other terms at $s=0$, i.e., $\kex_0$, $\kms_0$ $\kOe_0$ or $\kST$.

The only remaining unknown parameter in Eqs.~(\ref{eq:dfG})~\&~(\ref{eq:dfD}) is the angular frequency $\omega(s)$. Fortunately it can simply be extracted from micromagnetic simulations having reached steady-state regime, as $\omega(s)=2\pi f(s)$. Thus, one can easily model $G(s)$ and $D(s)$ by performing a fit on several simulation results. Let us precise that generally, the input variable of micromagnetic solvers is the injected current. We then obtain $\omega(\Jdc)$ as well as $s(\Jdc)$, involving a supplementary substitution.

Micromagnetic simulations are performed using mumax$^3$, a GPU-based program \cite{vansteenkiste2014design}, following the same protocol as for our previous study \cite{araujo2022ampere}. A free layer made of permalloy, with a radius $R$ of 100~nm and a thickness $h$ of 10~nm is considered. The used material parameters are presented in Table~\ref{tab:material}. The magnetic dot is discretized into $2.5\times 2.5$~nm$^2$ cells and two layers with a thickness of 5~nm each. Current densities ranging from 0 to 10~MA/cm$^2$ are injected into the junction, in the positive $z$-direction. The vortex polarity $P$ is thus fixed at -1, to respect the condition for stable auto-oscillations, i.e., $\Jdc Pp_z<0$.

\begin{table}[ht]
    \centering
    \caption{Material parameters of the magnetic tunnel junctions considered in the micromagnetic simulations and the analytical model.}
    \begin{tabular}{c c c c}
    \hline \hline
        Parameter & Symbol & Value & Units\\ \hline
        \makecell{Saturation \\magnetization} & $\Ms=\Ms^\text{ref}$ & 800 & emu/cm$^3$\\
        \makecell{Exchange stiffness\\coefficient} & $A$ & 1.07$\cdot 10^{-6}$ & erg/cm\\ 
        \makecell{Gilbert damping\\constant} & $\alpha_\text{G}$ & 0.01 & -\\ 
        \makecell{Spin-current\\polarization} & $p_J$ & 0.2 & -\\ \hline \hline
    \end{tabular}
    \label{tab:material}
\end{table}
Suitable physical simulation time is chosen to reach steady-state regime for each current. The reduced orbit radius $s$ is internally computed by mumax$^3$, while the frequency is retrieved thanks to our SNIFA technique \cite{SNIFA}. Magneto-crystalline anisotropy and the effect of temperature are neglected in this study. Three chiral configurations are examined, depicting the impact of the Ampère-Oersted field (AOF) $\HOe$ on the STVO dynamics: one without taking $\HOe$ into account, one with the planar vortex magnetization parallel to $\HOe$ ($C=+1$) and one anti-parallel to $\HOe$ ($C=-1$). Those will be labelled as noOe, $C^+$ and $C^-$, respectively.

\section{Results \& Discussion}
\begin{figure}[ht]
    \centering
    \includegraphics[scale=1]{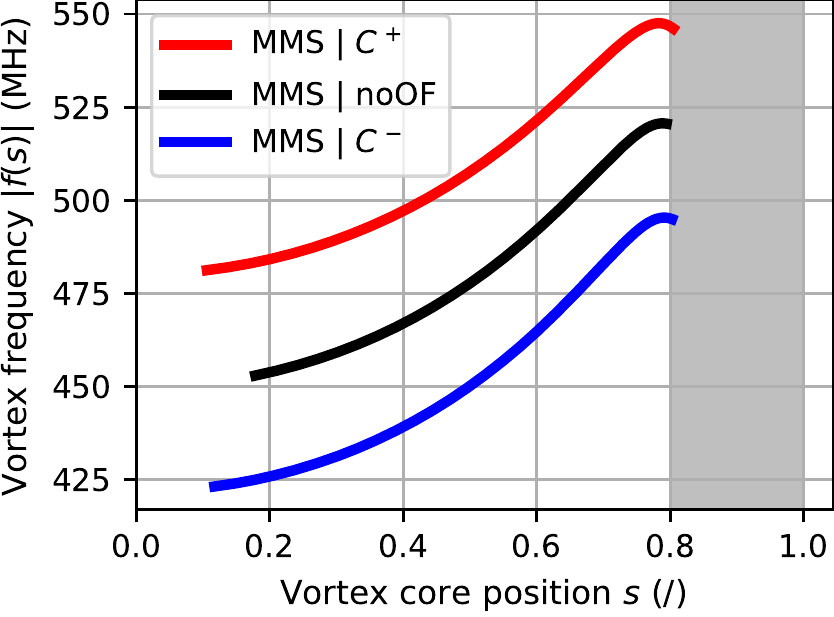}
    \caption{Absolute vortex gyrotropic frequency $|f(s)|$ as a function of the reduced vortex core position $s$, obtained from micromagnetic simulations (MMS). The colours black, red and blue correspond to simulations without AOF (noOF), with AOF and $C = +1$ ($C^+$) and with AOF and $C = -1$ ($C^-$), respectively. Only data in the steady-state oscillating regime are shown here, i.e., $s \in ]0,0.8]$. Cubic interpolations of the data points are depicted here with the respective color lines. The grey area represents the limit of vortex stability.}
    \label{fig:f(s)}
\end{figure}

The intrinsic property of our STVO, namely its gyrotropic frequency $f(s)=\omega(s)/(2\pi)$ with respect to the reduced orbit radius $s$, is presented on Fig.~\ref{fig:f(s)}. Those micromagnetic simulation data correspond only to the steady-state oscillating regime, i.e., $s\in]0,0.8]$. To increase the number of data points available for the fit, we performed a cubic interpolation on simulation results. Obviously, different sets of fitting coefficients (see Eqs.~(\ref{eq:fG})~\&~(\ref{eq:fD})) are derived for each chiral configuration. Also, the fitting bounds are chosen such as to stay in the nonlinear steady-state oscillating regime. The frequency retrieved from mumax$^3$ was originally expressed as $\omega(\Jdc)$. However, as we also obtained the evolution of $s(\Jdc)$ with respect to the current density \cite{araujo2022ampere}, we can easily express Eqs.~(\ref{eq:dfG})~\&~(\ref{eq:dfD}) as either being $s$ or $\Jdc$ dependent. Data points starting from 5.8, 6.2, 6.5~MA/cm$^2$ and ending at 8.4, 9, 9.8~MA/cm$^2$ were examined for $C^-$, noOF and $C^+$, respectively. Those current densities range thus between the first and second corresponding critical currents $J_\text{c1}$ and $J_\text{c2}$ \cite{araujo2022ampere}.

Since the evolution of the angular frequency $\omega(s)$ is known thanks to the simulations, the right hand side of Eq.~(\ref{eq:dfG}) can be determined for any value of $s$. A nonlinear least squares method is then used to fit the polynomial function proposed in Eq.~(\ref{eq:fG}). The obtained coefficients are reported in Table~\ref{tab:fG}. As stated previously a global correction factor $\delG$ is also introduced to take into account any discrepancies between the fully analytical TEA model and the MMS results, even in the resonant regime (i.e., $s=0$). One can observe that $\delG > 1$ for the three configurations which could either indicate an underestimation of the gyroconstant $G_0$ or an overestimation of the stiffness parameters term $\kex + \kms + C\Jdc \kOe$.
\begin{table}[ht]
    \centering
    \caption{Coefficients of the polynomial function given in Eq.~(\ref{eq:fG}), for each chiral configuration. Those were calculated from Eq.~(\ref{eq:dfG}) by a least squares nonlinear fit, after a cubic interpolation on the micromagnetic simulation data as presented on Fig.~\ref{fig:f(s)}. The maximum relative error between fit and simulation is also indicated.}
    \begin{tabular}{c c c c c c c}
    \hline \hline
         & $\delG$ & $\aG$ & $\bG$ & $\cG$ & $\dG$ & max($|$err$|$) [\%]\\ \hline
         $\Cp$ & 1.0904 & -0.0347 & 0.5118 & -1.5998 & 1.8183 & 0.328\\ 
         \noOe & 1.0977 & -0.0504 & 0.5799 & -1.7092 & 1.8900 & 0.241\\ 
         $\Cm$ & 1.1054 & -0.0417 & 0.5835 & -1.8082 & 2.0513 & 0.355\\ \hline \hline
    \end{tabular}
    \label{tab:fG}
\end{table}

\begin{figure}[ht]
    \centering
    \includegraphics[scale=1]{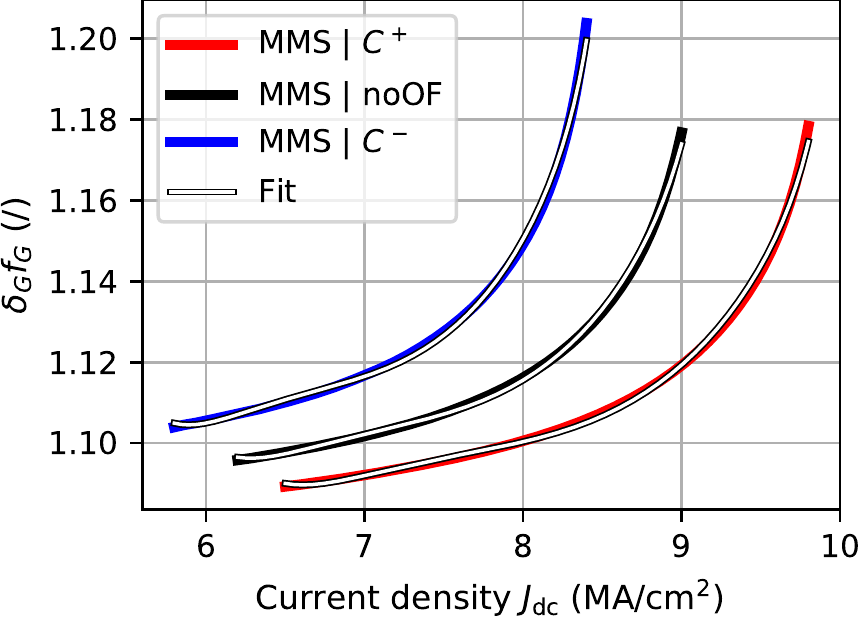}
    \caption{Value of $\delG\fG(s)$ as a function of the input dc current density $\Jdc$ for each chiral configuration $C^+$, noOF and $C^-$, identified in red, black and blue colors, respectively. The results calculated from Eq.~(\ref{eq:dfG}) (colored lines) were obtained after a cubic interpolation performed on the micromagnetic simulation (MMS) data presented on Fig.~(\ref{fig:f(s)}). The corresponding polynomial fits (see Eq.~(\ref{eq:fG})) are represented with white lines.} 
    \label{fig:G(J)}
\end{figure}

The evolution of the $\delG\fG(s)$ function with respect to the input dc current density $\Jdc$ is depicted on Fig.~\ref{fig:G(J)}. Continuously increasing nonlinear curves are obtained. The compensation of the exchange component could especially become dominant for high currents, as it diverges for value of $s\rightarrow 1$ (see Eq.~(\ref{eq:kex})). A strong splitting between the three configurations is noticed. This observation, which will be the subject of a future communication, could indicate that the chirality has an influence on the potential magnetic energy terms. As demonstrated in previous works \cite{khvalkovskiy2009vortex, araujo2022ampere,abreu2016controlling,choi2009quantitative}, the Zeeman term sign is directly impacted by the direction of the Ampère-Oersted field. However such splitting effect is not straightforward concerning the exchange and magneto-static terms.

An identical protocol is used to extract the $s$-dependence of the damping term. Knowing $\omega(s)$, the right hand side of Eq.~(\ref{eq:dfD}) can be calculated. Once fitted to the polynomial function given in Eq.~(\ref{eq:fD}), the coefficients reported in Table~\ref{tab:fD} are obtained. As for the gyrotropic term, one can observe that the global correction factor $\delD>1$, reporting either a too important $\Jdc \kST$ value or an underestimated damping constant $D_0$ when the vortex core is at the center of the dot.
\begin{table}[ht]
    \centering
    \caption{Coefficients of the polynomial function given in Eq.~(\ref{eq:fD}), for each chiral configuration. Those were calculated from Eq.~(\ref{eq:dfD}) by a least squares nonlinear fit, after a cubic interpolation on the micromagnetic simulation data as presented on Fig.~\ref{fig:f(s)}. The maximum relative error between fit and simulation is also indicated.}
    \begin{tabular}{c c c c c c c}
    \hline\hline
         & $\delD$ & $\aD$ & $\bD$ & $\cD$ & $\dD$ & max($|$err$|$) [\%]\\ \hline
         $\Cp$ & 1.0267 & 0.3499 & 0.1220 & 0.1790 & 0.0494 & 0.058\\
         \noOe & 1.0345 & 0.3017 & 0.1112 & 0.1301 & 0.0513 & 0.026\\
         $\Cm$ & 1.0423 & 0.2629 & 0.0793 & 0.1412 & 0.0168 & 0.030\\ \hline\hline
    \end{tabular}
    \label{tab:fD}
\end{table}

\begin{figure}[ht]
    \centering
    \includegraphics[scale=1]{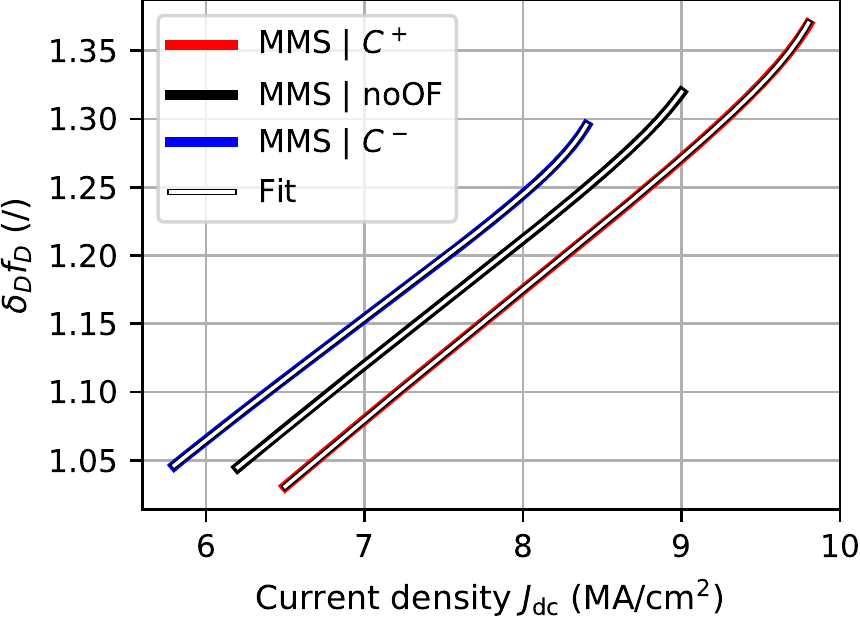}
    \caption{Value of $\delD\fD(s)$ as a function of the input dc current density $\Jdc$ for each chiral configuration $C^+$, noOF and $C^-$, identified in red, black and blue colors, respectively. The results calculated from Eq.~(\ref{eq:dfD}) (colored lines) were obtained after a cubic interpolation performed on the micromagnetic simulation (MMS) data presented on Fig.~(\ref{fig:f(s)}). The corresponding polynomial fits (see Eq.~(\ref{eq:fD})) are represented with white lines.}
    \label{fig:D(J)}
\end{figure}

The value of $\delD\fD(s)$ as a function of $\Jdc$ is given on Fig.~\ref{fig:D(J)}. A much more linear evolution is reported, compared to $\delG\fG(s)$ (see Fig.~\ref{fig:G(J)}). Again, a clear splitting between the three configurations is noticed. As it was the case for $\delG\fG(s)$, the respective distances between the $C^+$ and $C^-$ cases and the noOF curve is different which confirm the strong nonlinear behavior of the STVO dynamics.

\begin{figure}
    \centering%
    \includegraphics[scale=1]{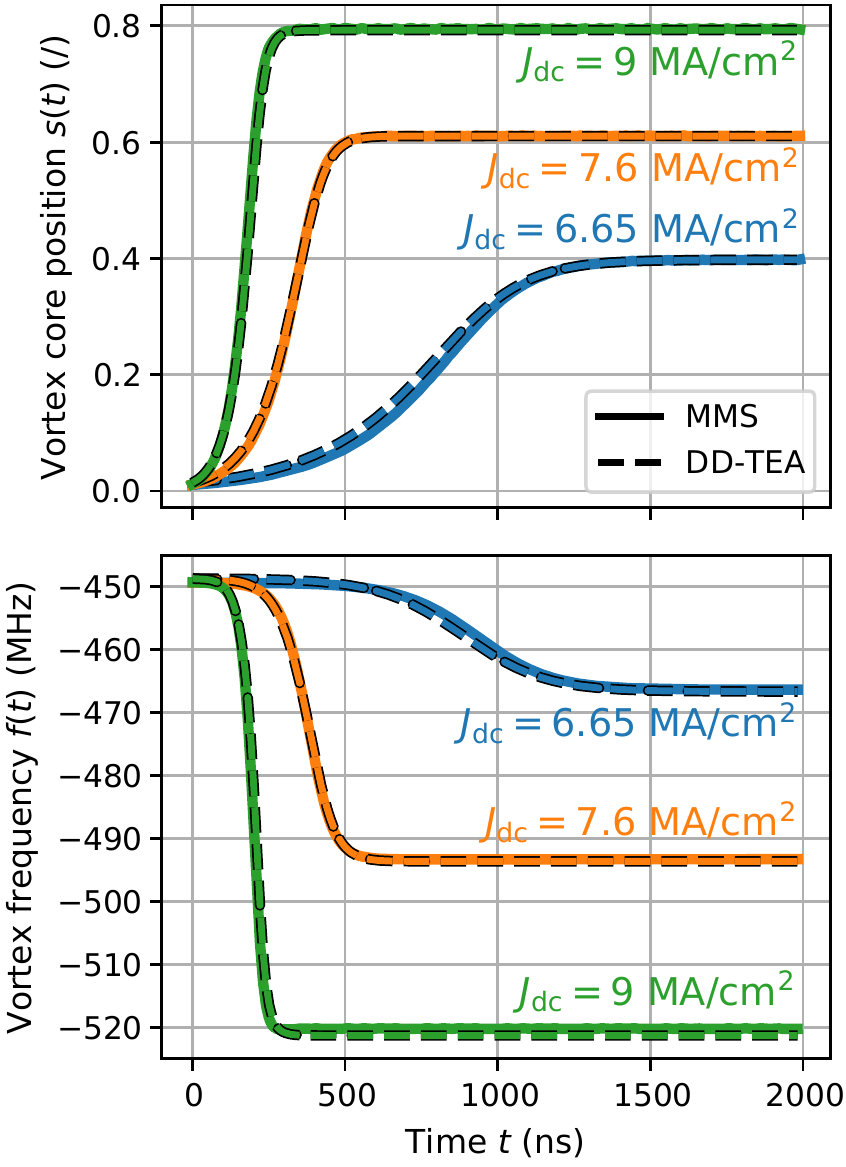}
    \caption{Transient regime calculations using micromagnetic simulations (MMS - continuous lines) and data-driven Thiele equation approach (DD-TEA - dashed lines) for (top) the reduced vortex core position $s(t)$ and (bottom) the vortex oscillation frequency $f(t)$. The input dc current densities $\Jdc$ applied are 6.65, 7.6 and 9 MA/cm$^2$ in blue, orange and green, respectively. The Ampère-Oersted field was not taken into account. All simulations were started at $s=0.01$.}
    \label{fig:f(t)}
\end{figure}

As data-driven corrections have been made to the TEA model (i.e., $G(s)$ and $D(s)$), we can now use it to predict the dynamics of our STVO with new input dc current density values by simply solving the system given in Eq.~(\ref{eq:matrix}). This novel hybrid method, which is semi-analytical, will be referred to as the data-driven Thiele equation approach (DD-TEA). A comparison between mumax$^3$ simulations and our DD-TEA model for the reduced vortex core position $s(t)$ and instantaneous frequency $f(t)$ with respect to time is given on Fig.~\ref{fig:f(t)}. The input currents were chosen to be between the two critical currents $J_\text{c1}$ and $J_\text{c2}$. The steady-state core position values are completely overlapped for both methods, even for  $\Jdc=9$ MA/cm$^2$ which leads to $s\approx0.8$, i.e., the limit of the model validity (but also the limit corresponding to the vortex polarity switch). Regarding the frequency, all curves start at $f\approx -449$~MHz. This frequency correspond to a vortex core position of $s=0.01$, the starting point of the simulations. As this orbit is close to $s=0$, the frequency is almost equal to $\omega_0 / (2\pi)$ \cite{araujo2022ampere}. With time increasing, the frequency evolves to the steady-state value corresponding to the current imposed. Various works \cite{gaididei2010magnetic, dussaux2012field,choi2009quantitative} propose TEA-based predictions of the steady-state orbit of oscillation and/or the frequency but rarely with such quantitative agreement, due to the lack of consideration of high-order terms. As far as the transient regime is concerned, only Guslienko {\it et al.} proposed a satisfactory analytical equation \cite{guslienko2014nonlinear}, to the best of our knowledge. In the present work, perfect correspondence is obtained. Relaxation time are consistent with simulations. This is not straightforward as only steady-state results were used to calibrate the model. Nothing but negligible shift between both methods is remarked in the transition for the lowest current curves.

Numerically solving the STVO dynamics using mumax$^3$ for 2000~ns, as presented in Fig.~\ref{fig:f(t)}, takes no less than three hours with the most powerful hardware to date (NVIDIA Tesla A100 GPU). Using now our DD-TEA model this calculation time is reduced to about 13~ms, an acceleration of more than 700k times. This number overcomes largely the already great 200 factor recently brought up by Chen {\it et al.} \cite{chen2022forecasting} for solving similar problems using artificial intelligence techniques. This remarkable speed-up can be explained by a dramatic reduction of the number of equations to be solved when using the DD-TEA instead of micromagnetic simulations. In mumax$^3$, the Landau-Lifshitz-Gilbert-Slonczewski \cite{landau1992theory,gilbert2004phenomenological,slonczewski1996current} equation is calculated in each cell of the magnetic structure, for each time step in the three spatial directions. For the specific dimensions used in this study, it corresponds to numerically solve 30144 equations between each timestep. Regarding the DD-TEA, only two equations are required, one for each in-plane coordinate of the vortex core, as the latter is seen as a quasi-particle. Furthermore, we believe that the speed-up factor obtained in this work would be further increased if our DD-TEA homemade code, for now running on a single CPU core, could be optimized to run on highly parallel GPU hardware. In addition, for larger radius dots even better numbers are expected as the number of cells scales up proportionally to $R^2$, for the same unit cell dimensions. Preliminary tests performed on a nanodot presenting a radius $R$ of 500~nm and a thickness $h$ of 9~nm (with same planar unit cell dimensions than for the present study) have given a speed-up factor of more than 2M times.

\section{Conclusion}
The dynamics of spin-torque vortex oscillators under out-of-plane input dc currents has been investigated. Starting from a fully analytical Thiele equation approach model developed previously \cite{araujo2022ampere}, data-driven corrections were brought to the gyrotropic and damping terms. These adjustments were obtained by fitting the model predictions to micromagnetic results performed using mumax$^3$. This hybrid method was then used to compute the STVO dynamics for new input current values and compared to micromagnetic predictions. An unprecedented agreement between both methods has been shown in the steady-state as well as in the transient regime. In addition to its accuracy, the DD-TEA model is faster than mumax$^3$ by a factor of about 1 million although the micromagnetic simulations were performed using the most powerful hardware available to date (NVIDIA Tesla A100 GPGPUs). Up to now, simulating STVO response for long-duration input signals was impractical as no method existed for performing both fast and precise simulations. The results presented in this paper fulfill both requirements and open the way for pioneering functionalization of such oscillators, namely in the framework of neuromorphic computing applications.


%

\section{Acknowledgements}
Computational resources have been provided by the Consortium des Équipements de Calcul Intensif (CÉCI), funded by the Fonds de la Recherche Scientifique de Belgique (F.R.S.-FNRS) under Grant No. 2.5020.11 and by the Walloon Region. F.A.A. is a Research Associate of the F.R.S.-FNRS. S.d.W. aknowledges the Walloon Region and UCLouvain for FSR financial support.

\section{Author contributions statement}
The study was designed by F.A.A. who created the analytical model. F.A.A. designed the micromagnetic simulations performed by S.d.W. and C.C.. S.d.W. wrote the core of the manuscript and all the co-authors (F.A.A., S.d.W. and C.C.) contributed to the text as well as to the analysis of the results.

\section{Data Availability}
The datasets generated during and/or analyzed during the current study are available from the corresponding author on reasonable request.

\section{Additional information}
The authors declare no competing interests.

\end{document}